\documentclass{elsart}
\usepackage[dvips]{graphicx}
\usepackage{amssymb}

\begin{document}

\begin{frontmatter}

\title{Scaling analysis of multivariate intermittent time series}
\author[ioc]{Robert Kitt\corauthref{cor1}}
\ead{kitt@ioc.ee}
\author[ioc]{Jaan Kalda},
\corauth[cor1]{Corresponding author. Tel.: +372 6204174; fax: +372 6204151}
\address[ioc]{Department of Mechanics and Applied Mathematics, Institute of Cybernetics at Tallinn University of Technology, 12061, Tallinn, ESTONIA}

\begin{abstract}
The scaling properties of the time series of 
asset prices and trading volumes of stock markets are
analysed. It is shown that similarly to the asset prices, 
the trading volume data obey multi-scaling length-distribution 
of low-variability periods. In the case of asset prices, 
such scaling behaviour can be used 
for risk forecasts: the  probability of observing next day  a 
large price movement is (super-universally) inversely proportional 
to the length of the ongoing low-variability period.
Finally, a method is devised for a multi-factor scaling analysis.
We apply the simplest, two-factor model to equity index and trading volume time series. 

\end{abstract}

\end{frontmatter}

\section{Introduction}
\label{intro}

Predicting future developments of financial asset prices
is an extremely difficult task. Certain problems, such as 
predicting the direction of movements, are nearly unsolvable.
Meanwhile, certain aspects can be easily predicted. 
For instance, one can be sure that the prices will always
fluctuate intermittently, largely due to people who believe they have
found the winning algorithm, the philosopher's stone.

The analysis of the historic charts of price dynamics is referred to as
technical analysis \cite{murphy}. The essence of the technical analysis is
the hypothesis that the patterns of historic data can forecast the future
price movements. On the other hand, the efficient market hypothesis (c.f.~\cite{fama}) 
states that security prices reflect fully all available information. 
In its century-long history dated back to the work of Bachelier
\cite{bach}, the financial analysis has made use of the both approaches.
The random walk hypothesis, standard deviation and the correlations
between securities' returns are the cornerstones of seminal papers in financial analysis \cite{markowitz,sharpe,black}. Econophysics has introduced various more 
elaborate, mostly non-linear tools of analysis (c.f.~\cite{stanley0,voit} and references therein).
Here we extend a recently developed technique of length-distribution of low-variability periods \cite{kitt}.

Strongly non-linear systems, such as turbulent fluids and plasmas, granular
media, biological and economical systems, etc., are typically characterized
by scale-invariance and scaling laws. So, it should be not surprising that
such seemingly different disciplines like turbulence studies, biological
physics, and econophysics have many common tools of data analysis. For
example, power-law distributions were first used in economics \cite{pareto}
in the end of 19th century, and later found in a wide variety of systems 
(often referred to as the Zipf's law), c.f.~\cite{zipf,ausloos2,gabaix,fujiwara}. 
Similarly, diverse systems are known to generate signals, which are 
self-affine, and hence, can be characterized by the Hurst exponent, 
c.f.~\cite{mandelFGN}. Further, the stable L\'evy distributions 
(c.f.~\cite{levy}) have been found to be relevant to all the mention
disciplines; about truncated Levy flights in econophysics, 
c.f.~\cite{stanley}. Finally, multi-fractal formalism is by far the most
popular tool for scale-invariant analysis of intermittent time-series, c.f.~\cite{mandel,vande}. 
It should be noted, however, that in many cases 
(including econophysical applications), there is no profound understanding of
the origin of multi-fractality, and hence, the  multi-fractal analysis is
not necessarily the optimal method. Indeed, multi-fractal formalism has
been devised in the context of turbulence, and is specifically suited for
systems with random multiplicative cascades \cite{multif}. Meanwhile, in
the case of such time-series as stock prices or heart rate variability
signal, the presence of multiplicative cascades is not evident.

Therefore, {\em there is a clear need for a deeper understanding of the character 
of intermittency in the case of financial time-series}. This problem can 
be approached by studying new independent and/or more general methods of 
scale-invariant analysis. For instance, in the case of heart rate variability, 
it has been found that the whole time-period can be clusterised into self-similarly 
distributed segments of approximately constant heart rate \cite{Galvan}. 
In order to address this problem, a new method has been suggested recently \cite{kitt} 
which is based on the analysis of the distribution of low-variability periods. 
The low-variability period is defined as a time period with maximal length where 
consecutive relative changes in realizations of the time series are less than given 
threshold $\delta$.  Note that in addition to the financial assets, the low-variability 
period analysis has been applied to biological systems \cite{kalda,kalda2}.
It has been shown \cite{kitt} that
in the case of the multi-affine time series, the cumulative distribution 
function of the low-variability periods is in the form of a
(multi-scaling) power-law. Since the opposite is not necessarily true, 
the low-variability period analysis is, indeed, a more universal method than 
the multi-affine analysis.

Even if the time-series is actually multi-affine, the low-variability period 
analysis can be still useful, because {\em (a)} the power-law exponent is related to 
the multi-fractal dimension; this circumstance provides an easy method for 
checking the assumption of multi-affinity; {\em (b)} low-variability periods 
provide higher time-resolution of time series analysis \cite{kitt}.

This paper serves three main purposes. {\em First}, we are going to apply the method of the 
analysis of low-variability periods to the data of trading volumes. This analysis is motivated as follows. Similarly to the asset prices, the trading volumes are known to fluctuate 
intermittently, c.f.~\cite{gopi,ausloos}. According to the Mandelbrot's model
of stock prices as a fractional Brownian motion in multi-affine trading time \cite{mandel}, one
could expect that the time series of trading volumes are multi-affine. The analysis of the length-distribution of low-variability periods of trading volumes serves as a test of this model. {\em Second}, we discuss the consequences of the presence of power laws of low-variability periods. 
{\em Third}, an attempt is made to generalize the method to multivariate time-series (e.g. a stock price together with the trading volume). It should be noted that in the case of multi-affine analysis, there is no simply interpretable way for such a generalization.

\section{Scaling of trading volumes}
\label{volume-calc}

We start with a brief description of the method devised in Ref.~\cite{kitt,kalda},
A low-variability period is defined as such a contiguous time-interval 
$T_i=[t_i,t_i+l_i]$,  which satisfies the conditions
\begin{equation} \label{delta}
|V(t)/\left<V(t)\right>_\tau-1| \le  \delta \;\;  \mbox{for} \;\; t \in T_i,
\end{equation}
and $|V(t)/\left<V(t)\right>_\tau-1| >  \delta$ for $t=t_{i}-1,t_{i}+l_i+1$;
it is assumed that the data sampling interval serves as the time unit.
The average of the trading volumes $V(t)$ is taken over a time window of 
length $\tau$:
\begin{equation}  \label{aver}
\left<V(t)\right>_{\tau}=\frac{1}{\tau}\sum_{k=0}^{\tau-1}V(t-k).
\end{equation}
Therefore, we have two control parameters: {\em (i)}
$\tau$ --- the length of the averaging window, and 
{\em (ii)} $\delta$ --- the variability threshold. 
Further, the cumulative length-distribution function of low-variability periods $R(n)$ is introduced:
$R(n)$ is the number of the low-variability periods of length $l_i\ge n$. 
We speak about multi-scaling behaviour, if the following power-law is observed:
\begin{equation} \label{power}
R(n)=R_0n^{-\alpha(\delta,\tau)},
\end{equation}
where $\alpha(\delta,\tau)$ is a scaling exponent and $R_0$ is a constant. 

It has been shown \cite{kitt} that in the case of multi-affine time-series,
the low-variability periods do follow a multi-scaling distribution law:
\begin{equation} \label{alphalog}
\alpha(\delta, \tau) = f(\log_{\tau}  \delta),
\end{equation}
where $f(h)$ denotes the  H\"older multifractal spectrum of the local Hurst exponents $h$.
Meanwhile, the presence of a scaling law (\ref{power}) does not necessarily imply multifractality.
Indeed, note that  the H\"older exponent $f(h)$ cannot exceed the topological dimensionality (one),
therefore the values $\alpha > 1$ are not related to multifractality. It should be stressed that
unequality $\alpha(\delta,\tau) > 1$ does neither imply the lack of multifractality:
in the case of multi-fractal time-series, $\alpha(\delta,\tau) > 1$ can (in principle) be observed for 
$\log_{\tau}  \delta < h_0$, where $h_0$ is the dominant local Hurst exponent, $f(h_0)=1$. For this range of 
parameters, the unequality $\alpha(\delta,\tau) > 1$ must be satisfied 
(assuming that $\alpha$ is a monotonous function of $\delta$).
Finally, if there is no data collapse $\alpha(\delta, \tau) \equiv f(\log_{\tau}  \delta)$
for $\alpha(\delta,\tau) \le 1$, the underlying time-series is certainly non-multifractal.
In the case of currency rates, a reasonable data collapse [according to Eq.~(\ref{alphalog})] has been observed.
In this Section, the daily trading volume data of various stock indices are tested for multi-scaling 
and multi-fractality.

\subsection{The data and the calculations}
\label{volume-data}

The stock market volumes are measured in the amount of shares traded on the exchange. The data used in the
analysis represent the daily closing prices and trading volumes and it is described in detail in the
Table \ref{data}.

\begin{table}[loc=htbp]
\caption{The data used in volume analysis}
\label{data}

\begin{tabular}{|l|p{5.5cm}|c|r|}

\hline
Abbr&Description&Calendar Period&\# of data\\ \hline
SPX&Standard \& Poor's 500 Index&04/01/93 - 13/09/04& 2947\\ \hline
DAX&The German Stock Index&04/01/93 - 13/09/04& 2950\\ \hline
NKY&Nikkei 225 Stock Average&04/01/93 - 14/09/04& 2879\\ \hline
MXEA&The MSCI Europe, Australasia and Far East Index&04/06/01 - 13/09/04& 832\\ \hline
CAC&CAC-40 Index of Paris Bourse&04/01/93 - 13/09/04& 2953\\ \hline
UKX&FTSE 100 Index&04/01/93 - 13/09/04& 2953\\ \hline
MXWO&MSCI World Index&04/01/01 - 13/09/04& 809\\ \hline
INDU&Dow Jones Industrial \-Average&04/01/93 - 13/09/04& 2941\\ \hline
TALSE&Tallinn Stock Exchange Index&25/02/03 - 13/09/04& 392\\ \hline
RTS&Russian Trading System Index&03/07/01 - 13/09/04& 794\\ \hline
WIG&Warsaw General Index&23/05/01 - 13/09/04& 836\\ \hline
BUX&Budapest Stock Exchange Index&15/09/97 - 13/09/04& 1747\\ \hline

\end{tabular}
\end{table}

If the cumulative distribution function $R(n)$ is plotted against $n$ in log-log scale, the power-law
Eq.~(\ref{power}) corresponds to a straight line. The scaling exponent $\alpha(\delta,\tau)$ was found 
as the slope of the linear trend line in log-log space, using the least-square fit method. 

The error of $\alpha(\delta,\tau)$ was estimated as follows: The least-squares fitted trend-line was found
as described above, except that the slope $\alpha$ was not optimised, i.e. it was considered as a fixed
parameter. Further, the sum of squared residuals $r(\alpha)$ was calculated as a function of $\alpha$.
The error estimate was found as $e = (\alpha'-\alpha)$, where $\alpha$ is the least-squares fitted value
of the slope, and $\alpha'$ satisfies the condition $r(\alpha') = 2r(\alpha)$.

\begin{figure}[loc=htbp]
\caption{Histogram of R-squared coefficients based on the regression analysis of determining 
scaling exponent $\alpha(\delta,\tau)$}
\includegraphics[width=13cm]{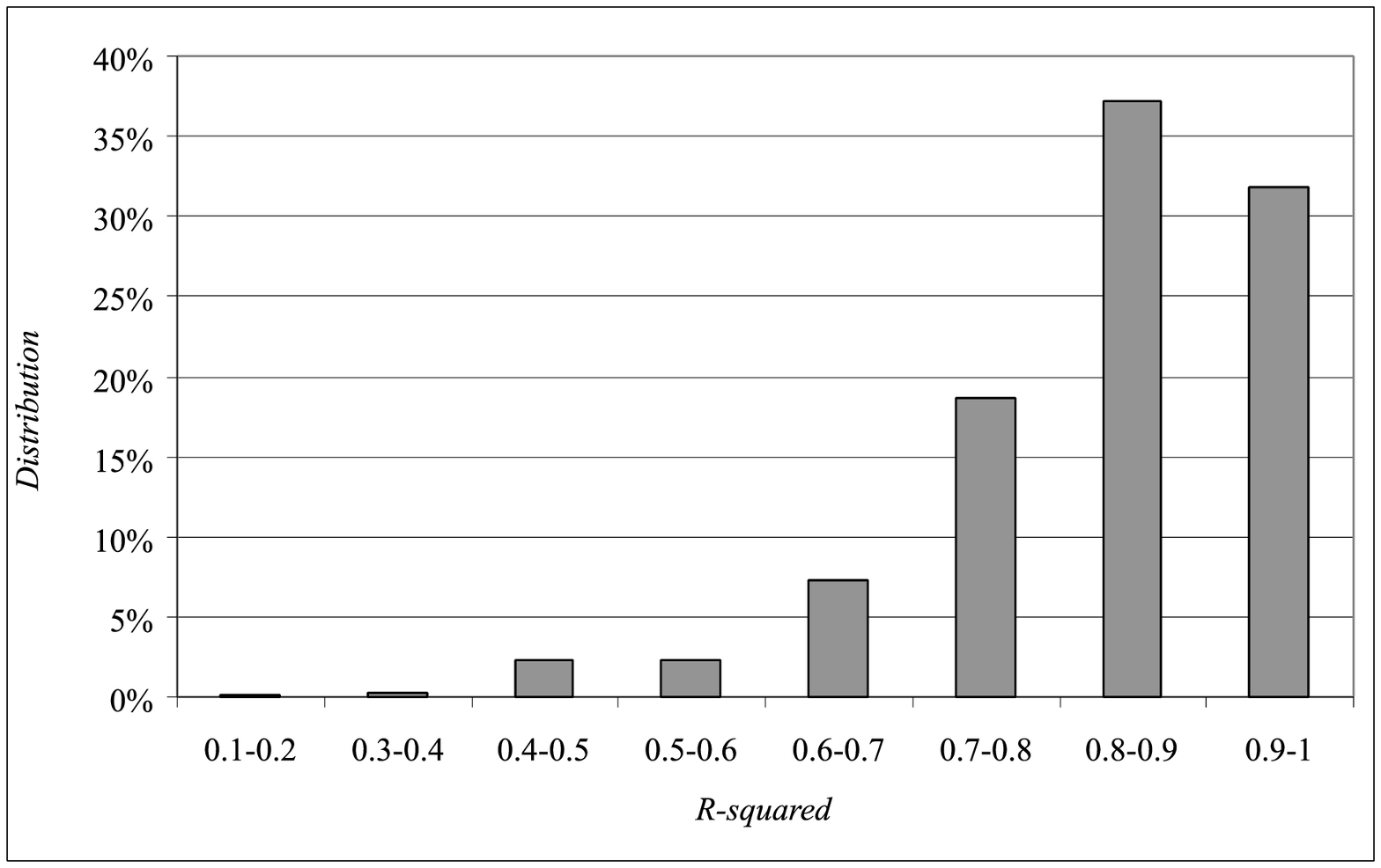}
\label{histo}
\end{figure}

We used $\delta=[5\%,10\%,...,70\%]$ and $\tau=[2,3,5,10]$ days as input parameters for all of the time
series described in Table \ref{data}. Total 650 calculations were carried out and in most cases the
least-square fit provided good results. The quality of the fit is measured by the $R^2$ coefficients that are plotted into the histogram in Fig \ref{histo}. The scaling exponents (found with $\tau=10$ days) with error estimates are presented in Table \ref{alphas}.

\begin{table}[loc=htbp]
\caption{The values of the scaling exponent $\alpha(\delta,\tau=10days$) in the case of daily trading volumes 
for various stock indices and different values of the parameter $\delta$}
\label{alphas}

\begin{tabular}{|r|c|c|c|c|c|c|}

\hline
$\delta$(\%)&CAC&DAX&INDU&NKY&SPX&UKX\\ \hline
5.0&3.49$\pm$0.45&3.71$\pm$0.33&3.39$\pm$0.40&3.16$\pm$0.30&3.55$\pm$0.46&3.65$\pm$0.39\\ 
10.0&3.34$\pm$0.47&3.21$\pm$0.35&2.63$\pm$0.33&3.01$\pm$0.43&2.75$\pm$0.32&2.94$\pm$0.28\\
15.0&2.89$\pm$0.39&2.61$\pm$0.28&2.25$\pm$0.19&2.78$\pm$0.42&2.13$\pm$0.26&2.73$\pm$0.27\\
20.0&2.63$\pm$0.27&2.44$\pm$0.37&1.90$\pm$0.15&2.07$\pm$0.27&1.95$\pm$0.23&2.50$\pm$0.30\\
25.0&2.44$\pm$0.23&2.33$\pm$0.37&1.79$\pm$0.22&2.00$\pm$0.18&1.68$\pm$0.14&1.97$\pm$0.24\\
30.0&2.34$\pm$0.26&1.71$\pm$0.23&1.69$\pm$0.24&1.70$\pm$0.16&1.37$\pm$0.15&1.83$\pm$0.19\\
35.0&1.99$\pm$0.23&1.58$\pm$0.26&1.65$\pm$0.19&1.36$\pm$0.13&1.14$\pm$0.12&1.52$\pm$0.15\\
40.0&1.66$\pm$0.19&1.62$\pm$0.20&1.24$\pm$0.14&1.49$\pm$0.18&1.10$\pm$0.10&1.38$\pm$0.17\\
45.0&1.78$\pm$0.20&1.42$\pm$0.17&1.12$\pm$0.16&1.39$\pm$0.21&0.98$\pm$0.11&1.05$\pm$0.17\\
50.0&1.55$\pm$0.20&1.44$\pm$0.20&1.44$\pm$0.13&1.67$\pm$0.19&0.82$\pm$0.09&1.20$\pm$0.13\\
55.0&1.35$\pm$0.16&1.32$\pm$0.19&1.21$\pm$0.11&1.56$\pm$0.17&0.80$\pm$0.09&0.96$\pm$0.11\\
60.0&1.41$\pm$0.20&1.22$\pm$0.15&1.11$\pm$0.12&1.47$\pm$0.16&0.55$\pm$0.08&0.94$\pm$0.12\\
65.0&1.15$\pm$0.15&1.01$\pm$0.15&0.98$\pm$0.13&1.42$\pm$0.13&0.77$\pm$0.08&0.86$\pm$0.13\\
70.0&1.37$\pm$0.19&0.91$\pm$0.14&0.98$\pm$0.12&1.25$\pm$0.10&0.71$\pm$0.07&0.89$\pm$0.14\\

\hline 
$\delta$(\%)&BUX&WIG&TALSE&RTSI&MXEA&MXWO\\ \hline
5.0&3.50$\pm$0.25&3.23$\pm$0.11&3.65$\pm$0.39&3.70$\pm$0.06&3.32$\pm$0.75&2.81$\pm$0.32\\ 
10.0&3.45$\pm$0.54&2.64$\pm$0.27&3.11$\pm$0.07&2.64$\pm$0.11&2.57$\pm$0.31&2.35$\pm$0.23\\
15.0&2.36$\pm$0.30&2.77$\pm$0.37&2.84$\pm$0.19&3.17$\pm$0.42&2.23$\pm$0.30&1.99$\pm$0.23\\
20.0&2.58$\pm$0.36&2.67$\pm$0.44&2.70$\pm$0.26&3.13$\pm$0.43&1.60$\pm$0.16&1.30$\pm$0.13\\
25.0&2.50$\pm$0.25&2.07$\pm$0.27&2.88$\pm$0.36&2.63$\pm$0.34&1.29$\pm$0.16&1.12$\pm$0.15\\
30.0&2.19$\pm$0.23&2.25$\pm$0.29&2.37$\pm$0.46&2.28$\pm$0.32&1.18$\pm$0.12&0.89$\pm$0.15\\
35.0&2.11$\pm$0.25&1.89$\pm$0.17&2.45$\pm$0.34&2.05$\pm$0.22&1.12$\pm$0.14&1.01$\pm$0.15\\
40.0&1.97$\pm$0.15&1.57$\pm$0.13&2.04$\pm$0.28&1.89$\pm$0.31&0.89$\pm$0.09&0.63$\pm$0.07\\
45.0&1.89$\pm$0.14&1.47$\pm$0.10&2.02$\pm$0.32&2.01$\pm$0.30&0.67$\pm$0.08&0.63$\pm$0.08\\
50.0&1.81$\pm$0.16&1.31$\pm$0.12&1.95$\pm$0.26&1.97$\pm$0.20&0.57$\pm$0.07&0.63$\pm$0.07\\
55.0&1.65$\pm$0.14&1.37$\pm$0.10&1.59$\pm$0.17&1.64$\pm$0.14&0.64$\pm$0.08&0.56$\pm$0.09\\
60.0&1.51$\pm$0.16&1.38$\pm$0.12&1.64$\pm$0.19&1.46$\pm$0.17&0.51$\pm$0.10&0.56$\pm$0.09\\
65.0&1.36$\pm$0.16&1.21$\pm$0.12&1.60$\pm$0.22&1.36$\pm$0.19&0.51$\pm$0.10&0.53$\pm$0.09\\
70.0&1.28$\pm$0.16&1.22$\pm$0.11&1.33$\pm$0.18&1.28$\pm$0.21&0.37$\pm$0.06&0.33$\pm$0.07\\ \hline

\end{tabular}
\end{table}

\subsection{Scaling properties of volume data and discussions}
\label{volume-scal}

\begin{figure}
\caption{Scaling exponent $\alpha(\delta,\tau)$ plotted against $log_{\tau}\delta$ for SPX and DAX time 
series for $\tau=3,5,10$ days}
\includegraphics[width=13cm]{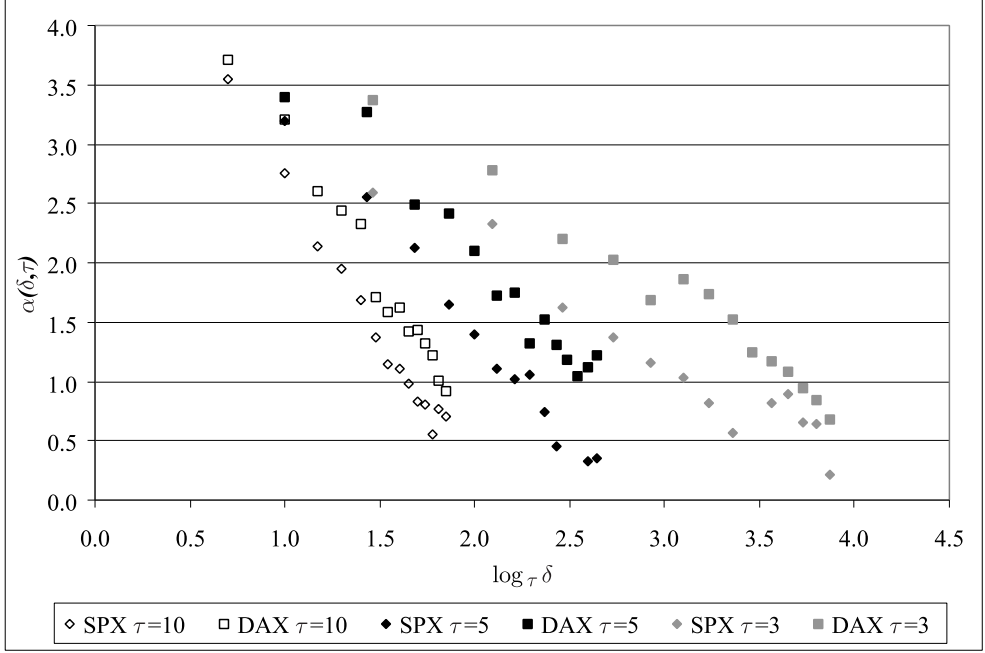}
\label{logtau}
\end{figure}

As seen from Fig.~\ref{histo} and Table \ref{alphas}, the low-variability periods of 
the daily volume time series follow reasonably well a multi-scaling behaviour
(similarly to what has been observed in the case of stock prices \cite{kitt}).
In Fig.~\ref{logtau}, the scaling exponent $\alpha(\delta,\tau)$ is plotted against the $log_{\tau}\delta$
for DAX and SPX time series with $\tau=3,5,10$ days. We can see that there is a departure from the 
multi-affine expectation
--- there is no data collapse at range $\alpha\le 1$ [which would have been corresponding to Eq.~(\ref{alphalog})]. 
This observation is interpreted as follows. Here,
each data point describes the amount of shares traded in respective stock exchange in a calendar
day. The condition $\alpha<1$ is satisfied only for very high thresholds ($\delta>50\%$). 
Typically, there are very few periods, when trading volumes would
fluctuate with such a large amplitude. Therefore, the number of valid data points is very low. 
As a consequence, the scaling range is narrow, and the formally calculated 
scaling exponent is non-stationary (depends on the particular realisation of the time-series).
This behaviour is different from what has been observed for currency time series (when
the results $\alpha \le 1$ were more stationary, with a reasonable data collapse 
in $\alpha-\log_\tau \delta$-plot).

So, we can conclude that while the trading volume data are intermittent and fluctuations are 
scale-invariant [described by Eq.~(\ref{power})], the degree of intermittency is lower than 
in the case of currency fluctuations: the scaling exponent values tend to be always larger than one.
Respectively, the multifractal pattern [like described by Eq.~(\ref{alphalog})] is not observable.

\subsection{Consequences of the scaling behaviour}
\label{consequences}
In the case of financial time-series, a crucial question is how to make forecasts. The most useful kind
of forecasts would give predictions of the direction of future prices. However, these forecasts will be always
very unreliable, the efficient market hypothesis denies such a possibility entirely (c.f.~second law of
thermodynamics). Meanwhile, the risk-related forecasts are completely possible (and may prove to be
useful). Therefore, a natural question arises: what are the risk-prediction-related consequences of the
presence of the power-law distribution of low-variability periods? More specifically, suppose we have had
a low-variability period, which had lasted $n$ days (and still goes on). Can we say something about the
possibility of today being the last day of this ``silent'' period? In other words, what is the
probability $p(n)$ that the tomorrow's movement exceeds our pre-fixed threshold $\delta$?

Apparently, this probability is given by the ratio of {\em (a)} the number of those low-variability periods, 
the length of which is exactly $n$, $N_a=R(n)-R(n+1)$, and {\em (b)} the number of those low-variability 
periods, which have length $m\ge n$, $N_b=R(n)$.
If $n$ is large, the difference
$R(n)-R(n+1)$ can be calculated approximately as $-\frac {dR}{dn}$. Upon applying Eq.~(\ref{power}) we arrive at $N_a\approx \alpha R_0 n^{-\alpha-1}$ and $N_b\approx R_0 n^{-\alpha}$. Bearing in mind that $p(n)=N_a/N_b$, the final result is written as \begin{equation} \label{silence}
p(n) \approx \alpha n^{-1}.
\end{equation}
 
Therefore, we have shown that the very presence of a power law of the low-variability periods has an
interesting consequence: the probability that the tomorrow's price movement will be larger than the
movements of $n$ preceding days, is inversely proportional to $n$. The predicted power law exponent is
independent of the scaling exponent $\alpha(\delta,\tau)$, i.e.\ we are dealing with super-universality. This 
super-universality appears to be related to the super-universality of the scaling of direct avalanches 
in self-organised critical systems \cite{maslov}. It allows us to make every day a series of forecasts: 
by scanning the values of the threshold parameter $\delta$, we find the length $n_\delta$ of the current 
(still ongoing) low-variability period. Then, the probability that the tomorrow's movement $p(\delta)$
remains below the threshold $\delta$ is estimated as
$p(\delta)\approx \alpha(\delta,\tau=2\,\mbox{days}) n_\delta^{-1}$. Note that the prefactor
$\alpha(\delta,\tau=2\,\mbox{days})\approx 1$ can be dropped, but will improve the forecast;
$\tau=2$ days means that only the prices of two subsequent days are compared.

\section{Two-factor model of scale invariance in financial time series}
\label{multi-model}

In the technical analysis, various forms of multi-factor models have been used for some time \cite{murphy}. In particular, it has been conjectured that in the case of 
trading volumes and price fluctuations, the higher-than average trading volumes generally ``confirm'' the price trend. However, according to our best knowledge, scale-invariant methods have not yet been
applied to the multi-factor studies.

The low-variability period analysis provides simply interpretable way of generalisation of multi-signal time series analysis. With the conventional multi-affine analysis, the interpretation of the results is significantly harder. Indeed, consider the wavelet transform $w(a,b)$, where $a$ denotes a coordinate and $b$ stands for the wavelet width. 
For a multi-affine signal, the following scaling law is expected: $\left< w(a,b)^q\right> \propto b^{\tau_q}$.  For a bi-variate analysis of two signals, 
this equation can be generalized as $\left< w_1(a,b)^q\cdot w_2(a,b)^p\right> \propto b^{\tau_{q,p}}$. Here, $w_1$ and $w_2$ denote the wavelet transform of the two 
signals. For two dependent signals,  $\left< w_1(a,b)^q\cdot w_2(a,b)^p\right>\ne \left< w_1(a,b)^q\right>\cdot \left< w_2(a,b)^p\right>$;
however, there is no clear way of interpreting the features of the scaling exponent $\tau_{q,p}$.
Therefore, in this section, the analysis of low-variability periods is generalised and the multi-factor model is proposed. In particular, we address the above mentioned statement about the relationship between trading volumes and stock prices. 

In our previous calculations, the low-variability periods are defined either via the condition (\ref{delta}), or via the same unequality, with trading volume $V(t)$ being substituted by stock price (or index value) $p(t)$ \cite{kitt}:

\begin{equation} \label{m0}
|p(t)/\left<p(t)\right>_\tau-1| \le  \delta_p  \;\; \mbox{for} \;\; t \in T_i.
\end{equation}
Hereinafter we refer to the application of Eq.~(\ref{m0}) as the usage of the {\em single-factor price model} or {\em Method 0} (for brevity). Similarly, the condition (\ref{delta}) corresponds to 
the {\em single-factor volume model} or {\em Method 0 (Volume)}. There are two
ways to generalize the concept of low-variability into multi-factor model:
\begin{enumerate}
\item Low-variability persists if both condition (\ref{delta}) and (\ref{m0}) are satisfied;
\item Low-variability persists if any of the conditions (\ref{delta}) and (\ref{m0}) are satisfied.
\end{enumerate}
It is clear that the number of low-variability periods is larger
by using the latter option. In fact, using the first option led us often to a very small number of low-variability periods 
(this, of course, is related to the limited length of the time-series). 
Therefore, here we present only the results corresponding to the second definition. 
Note that for multi-factor models with more than two inputs, the conditions similar to Eq.~(\ref{m0}) can be combined with any set of logical operators ``and'' and ``or''.

In what follows we use several definitions of the two-factor low-variability periods (the application of these definitions will be referred to as {\em Method 1 -- Method 3}, respectively):

\begin{equation} \label{m1}
(|p(t)/\left<p(t)\right>_\tau-1| \le  \delta_p) \bigvee (|V(t)/\left<V(t)\right>_\tau-1| \le  \delta_v) \;\; \mbox{for} \;\; t \in T_i.
\end{equation}

\begin{equation} \label{m2}
(|p(t)/\left<p(t)\right>_\tau-1| \le  \delta_p) \bigvee (V(t)/\left<V(t)\right>_\tau-1 \le  \delta_v) \;\; \mbox{for} \;\; t \in T_i.
\end{equation}

\begin{equation} \label{m3}
(|p(t)/\left<p(t)\right>_\tau-1| \le  \delta_p) \bigvee (V(t)/\left<V(t)\right>_\tau-1 \ge  -\delta_v) \;\; \mbox{for} \;\; t \in T_i.
\end{equation}

\subsection{Asymptotic behaviour}
\label{border}

Equation (\ref{m1}) is symmetric with respect to the price and volume conditions. The low-variability period is terminated
as soon as the relative price change exceeds (by modulus) $\delta_p$, and the relative volume change exceeds (by modulus) $\delta_v$. 

If the condition (\ref{m1}) is applied, the following asymptotic laws are expected:
\begin{displaymath}
\alpha(\delta_p,\delta_V,\tau) \to \alpha(\delta_p,\tau) \;\;\mbox{when} \;\; \delta_V \to 0,
\end{displaymath}
and
\begin{displaymath}
\alpha(\delta_p,\delta_V,\tau) \to \alpha(\delta_V,\tau) \;\;\mbox{when} \;\; \delta_p \to 0.
\end{displaymath}

\begin{figure}
\caption{The convergence of {\em Method 1} $\to$ {\em Method 0}. 
The scaling exponents $\alpha(\delta_p,\delta_V,\tau)$ are calculated for 
DAX time series by using conditions (\ref{m0}) and (\ref{m1})  with $\tau=10$ days 
and $\delta_v = [50\%,25\%,15\%,5\%,0\%]$}.
\includegraphics[width=13cm]{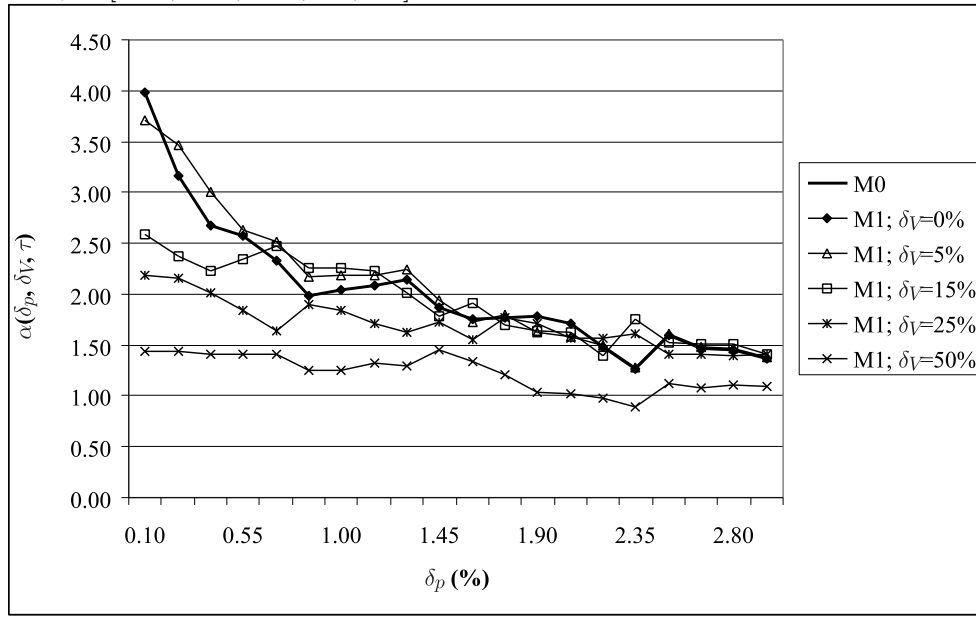}
\label{conv-m1DAX}
\end{figure}

\begin{figure}
\caption{The convergence of {\em Method 1} $\to$ {\em Method 0 (Volume)}. 
The scaling exponents $\alpha(\delta_p,\delta_V,\tau)$ are calculated for 
SPX time series by using conditions (\ref{delta}) and (\ref{m1}) with $\tau=10$ days and $\delta_p = [2.5\%,1.9\%,1.0\%,0.1\%,0.0\%]$}.
\includegraphics[width=13cm]{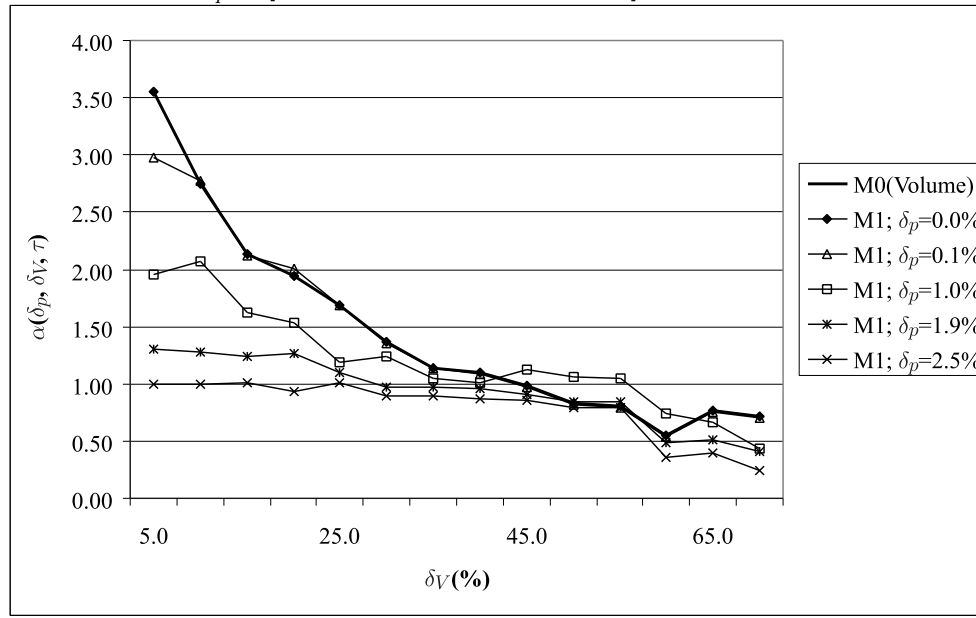}
\label{conv-m1SPX}
\end{figure}

Such a behaviour is, indeed,  observed for the time series of DAX and SPX. 
The data used in the calculations is described in Table \ref{data}. 
In Figures \ref{conv-m1DAX} and \ref{conv-m1SPX} the scaling exponents $\alpha(\delta_p,\delta_V,\tau)$ are 
displayed for DAX and SPX time series by using $\tau=10$ days. 
>From Fig.~\ref{conv-m1DAX}, it is seen that for low values of $\delta_V$, 
the low-variability condition for volume is almost always violated.  
Therefore, the scaling exponent is defined almost solely with the price  parameter
(the curves with $\delta_V=0$ and  $\delta_V=0.5$ are very close to each other). 
On the other hand, for high values of $\delta_V$, the volume condition is typically 
satisfied, and the horizontal lines (in $\alpha$-$\delta_p$ space) denote a low 
dependence on the price parameter. 
These arguments apply to the dependence of the exponent $\alpha(\delta_p,\delta_V,\tau)$ on the 
price threshold parameters, as well, see
Fig.~\ref{conv-m1SPX}. 

\subsection{Methods 2 and 3: differences between volume spike and volume squeeze}
\label{volume-diffs}
According to Eq.~(\ref{m2}), the low-variability period is terminated when price 
change (rise or drop) is significant (i.e.~larger than the threshold parameter), 
and the volume increases faster than the threshold. Equation (\ref{m3}) represents a definition, opposite to Eq.~(\ref{m2}): the low-variability period is terminated when the price change exceeds the 
threshold parameter, and the volume decreases faster than the threshold. These definitions are useful for studying the asymmetry between the volume rise and drop:
if the multi-scaling exponent $\alpha(\delta_p,\delta_V,\tau)$ turns out to be different for Methods 
2 and 3, there must be an asymmetry between those volume spike and
squeeze events, which are accompanied by a large price variability.
Indeed, the price condition in Eqns (\ref{m2}) and (\ref{m3}) is the same; so, the differences in the scaling exponent
$\alpha(\delta_p,\delta_V,\tau)$  must be due to the different effect of the volume condition (\ref{m2}) and (\ref{m3}).

Let us refer back to Method 1 [Eq.~(\ref{m1})]. With this method, 
the events terminating the low-variability periods represent a superposition of the respective events 
for the Methods 2 and 3 [Eqns (\ref{m2}) and (\ref{m3})].  
So, if the scaling exponents calculated
according to Method 2 are very similar to the ones calculated according to Method 1, we can
conclude, that the amount of the low-variability periods defined by Method 3 is insignificant. The
same holds true if the scaling exponents of Method 3 tend to be similar to the ones of Method
1.

\begin{figure}
\caption{Multi-affine scaling of the two-factor model. The scaling exponents $\alpha(\delta_p,\delta_V,\tau)$ 
of DAX time series are plotted against $log_{\tau=10}\delta_p$ with {\em (a)} $\delta_V=5\%$, {\em (b)} $\delta_V=15\%$, 
{\em (c)} $\delta_V=25\%$ and {\em (d)} $\delta_V=50\%$}
\includegraphics[width=14cm]{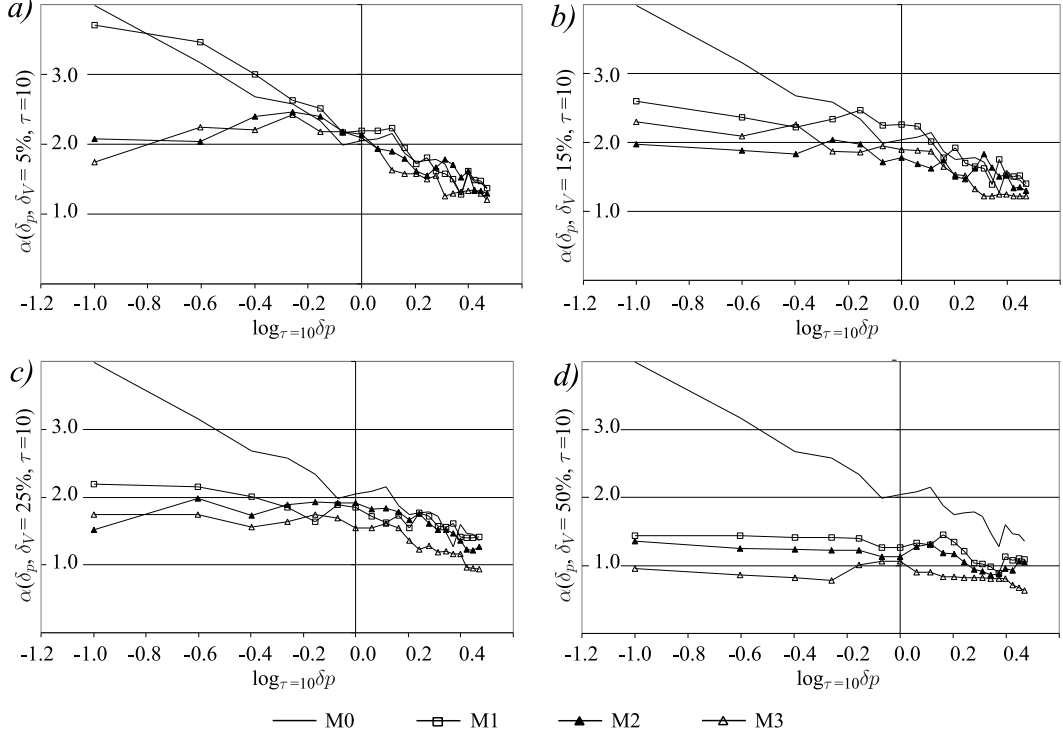}
\label{2f-logtau}
\end{figure}

The multi-factor scaling exponents $\alpha(\delta_p,\delta_V,\tau)$ are found for the DAX data with
$\tau=10$ days. In Fig.~\ref{2f-logtau}, the scaling exponents $\alpha(\delta_p,\delta_V,\tau=10)$ of DAX
index are plotted against $log_{\tau=10}\delta_p$ with {\em (a)} $\delta_V=5\%$, {\em (b) }$\delta_V=15\%$,
{\em (c)} $\delta_V=25\%$ and {\em (d)} $\delta_V=50\%$ . The Methods 0-3 [Eqns (\ref{m0})--(\ref{m3})] are used for definition of low-variability periods.

\subsection{Discussion of the results}

By giving a small value to the thresholds $\delta_V$ and $\delta_p$  in any of Eqns (\ref{m0})--(\ref{m3}), 
the number of longer low-variability 
periods becomes small. Therefore, we have a large number  of short low-variability periods 
(compared to high $\delta$ values); this leads to high values of the scaling exponent. 
In Fig.~\ref{2f-logtau}a, the threshold value of $\delta_V$ is set to very low 
level of $5\%$. From Eq.~(\ref{m1}) it can be seen that then the two-factor model leads to results which are very similar to  the single-factor (i.e.\ price factor) model. 
Likewise, the larger the volume threshold $\delta_V$, the larger the difference between the 
$\alpha$--$\delta_p$-curves of the single- and multi-factor models (at $\delta_V=50\%$, the 
curves are rather dissimilar).

An important issue is the difference between the curves of Methods 2 and 3.
One can notice that there is a different behaviour
at the small values of the parameter $\delta_p$, an evidence of the asymmetry between volume rise and drop.
One can also notice that the scaling exponents calculated according to the
Method 3 [Eq.~(\ref{m3})] are lower than the ones calculated according to Methods 1 and 2 [Eqns (\ref{m1}) and (\ref{m2})].
Meanwhile, the difference between the outcomes of the Methods 1 and 2 [Eqns (\ref{m1}) and (\ref{m2})] is minor.
Therefore, we conclude, that high price variability is typically accompanied by increasing volume.
This conclusion is independent of the price/volume pre-history (i.e. is valid both for short and 
long low-variability periods).

In this paper, we have not analysed the problem of higher-rank multi-factor models. 
However, this can be useful for e.g. multi-stock data analysis, where each stock price provides an 
independent input stream. This situation will be addressed in further studies. 

\section{Conclusion}
The concept of low-variability periods has been proven to be useful for various econophysical
issues (not just limited to the scope of stock prices/indices
and currency exchange rates). So, we found that the time series of stock
trading volumes obey multi-scaling properties, similarly like the price data. 
However, while the multi-scaling exponent of the price time series follows a 
pattern, characteristic to the multi-affine data, in the case of trading volumes, 
there is a clear departure from that pattern (one can say that the fluctuations are less intermittent).

Further, we have shown that the presence of the multi-scaling distribution of the 
length-distribution of the low-variability periods gives rise to a super-universal
scaling law for the probability of observing next day  a 
large price movement. This probability is inversely proportional 
to the length of the ongoing low-variability period, a fact which can be used for risk forecasts.

Finally, the multi-factor
model is proposed for time series analysis. In this paper, only the simplest two-factor model is
described and applied to stock price and volume data. The low-variability periods of multi-variate
time series can be defined in different ways; for instance, the threshold conditions applied to the 
single data streams can be combined by logical ``and'', as well as by logical ``or''.
In the our case of price and volume data, three different definitions of low-variability periods have been
applied (in order to study the asymmetry between the volume rise and drop, we have also applied 
sign-dependant threshold conditions). This analysis led us to the conclusion that 
high price variability is typically accompanied by increasing trade volume,
independently of the prior events of the market. In the light of this observation, the 
common thesis of technical analysis, ``increased trading volumes confirms the price trend'',
becomes less useful. Indeed, most of the significant price jumps are accompanied 
by increased trading volumes; so, almost all the ``price trends'' pretend to be ``confirmed''.

\section*{Acknowledgement}
The support of Estonian SF grant No.\ 5036 is acknowledged. We would also like to thank prof. J\"uri Engelbrecht for fruitful discussions.

%


\end{document}